\documentclass[prb,aps,superscriptaddress,twocolumn]{revtex4}%
\usepackage{graphicx}
\usepackage{dcolumn}
\usepackage{bm}
\usepackage{color}
\usepackage{ulem}
\usepackage{amsmath}
\usepackage{amsfonts}
\usepackage{amssymb}%
\providecommand{\U}[1]{\protect\rule{.1in}{.1in}}

\begin{document}
\title{The free energy of quantum spin systems: Functional integral representation}
\author{Peter W\"{o}lfle}
\affiliation{Institute for Condensed Matter Theory and Institute for Nanotechnology,
Karlsruhe Institute of Technology, D-76128 Karlsruhe, Germany}
\author{Natalia B. Perkins}
\affiliation{School of Physics and Astronomy, University of Minnesota, Minneapolis, MN
55116, USA}
\author{Yuriy Sizyuk}
\affiliation{School of Physics and Astronomy, University of Minnesota, Minneapolis, MN
55116, USA}

\begin{abstract}
In this work, we propose a method for calculating the free energy of
anisotropic quantum spin systems. We use the Hubbard-Stratonovich
transformation to express the partition function of a generic bilinear
super-exchange Hamiltonian in terms of a functional integral over classical
time-dependent fields. In the general case the result is presented as a product of traces over single spins subject to a time-dependent field. The traces may be evaluated in closed form in the case of Ising-type spin systems. In the General case we derive a compact expression for the contribution of
Gaussian spin fluctuations to the free energy. We show how anisotropic spin interactions lead to anisotropies in the free energy, giving rise to pinning of the spontaneous magnetization along preferred directions

\end{abstract}
\maketitle

\section{Introduction}

Recent research activities on 4$d$ and 5$d$ transition metal oxides have shown
that the interplay of the strong spin-orbit coupling (SOC), crystal field
interactions, and electron correlations may lead to strongly anisotropic,
bond-dependent exchange interactions between localized magnetic
moments.\cite{khaliullin05,jackeli09,jackeli10,perkins14,sizyuk14,rau14,kimchi14,nussinov15,balents14,
chaloupka15,chaloupka16,ioannis15,ioannis16,
trebst15,jackeli15} These anisotropic interactions have the generic form
$J_{j,j^{\prime}}^{\alpha\alpha^{\prime}}S_{j}^{\alpha}S_{j^{\prime}}%
^{\alpha^{\prime}}$, in which $S_{j}^{\alpha}$ denotes the $\alpha$-th
component of the pseudospin operator on site $j$, describing the magnetic
degree of freedom, $\mathbf{{\mathcal{J}}}_{\rm eff}$.  The most notable examples  appear in iridium and ruthenium systems,\cite{rau16}
in which  magnetic degrees of freedom on  Ir$^{4+}$ or Ru $^{3+}$ ions arise
 from a low-lying multiplet (usually a doublet) of total
angular momentum $\mathbf{{\mathcal{J}}}_{\rm eff}=\mathbf{{\mathcal{L}}}_{\rm eff}%
+\mathbf{{\mathcal{S}}}=1/2$,  in which  spin  and  orbital  angular  momenta  are  intertwined  due  to  the  strong  SOC.

In systems with strong SOC, the exchange
interactions $J_{j,j^{\prime}}^{\alpha{\alpha^{\prime}}}$ are generally not
SU(2) invariant and are lattice-specific, because magnetic moments contain
both spin and orbital angular momentum components. 
The presence of anisotropic bond-directional spin interactions in SOC
models provides the foundation for the realization of a plethora of novel
quantum ground states, such as the Kitaev spin liquid,\cite{kitaev06} and a
rich variety of magnetically ordered states\cite{choi12,chun15,banerjee2015,johnston15}  including single- and
multiple-$Q$ spirals.\cite{biffin14-1,biffin14-2,tomo15} 
While the theoretical
analysis of the structure of magnetic ground states of the SOC magnets has
recently received considerable attention, only few investigations have
addressed the problem of how the direction of the order parameter is
selected.\cite{balents14,chaloupka15,chaloupka16,ioannis15,ioannis16}
  
There are two possible scenarios  for the selection of the direction of the order parameter.
 In the most generic anisotropic biquadratic  models,  it   might be selected  already at the  mean field level by off-diagonal pseudospin interactions  even if  these interactions are subdominant with respect to diagonal couplings.\cite{sizyuk16,chaloupka16}   In the exchange models with only diagonal interactions,  i.e. in the  compass-like models,  the magnetic orders  with different directions of the order parameters are degenerate on the mean field  level.
 However,  this classical degeneracy is accidental and is therefore lifted by the order by
disorder mechanism  which selects a discrete set of  states,
each with a particular direction of the order parameter.

 The  thermal\cite{villain80}  and quantum order-by-disorder\cite{shender82} 
mechanisms selecting a particular spin ordering pattern from a  classically degenerate manifold of states have been proposed to be at play in a number of condensed matter systems.
\cite{ioannis15,chubukov92,zhitomirsky12,chern13,gingras14,chernyshev14,chernyshev15}
However, in  most of the cases the quantum fluctuations are considered only at zero temperature, while  at finite temperatures only classical spin fluctuations are considered .

The purpose of this paper is to present a general approach to compute the free
energy of quantum spin systems with anisotropic interactions and study how
spin fluctuations explicitly break the degeneracy at the mean-field level
and select a particular direction of the order parameter from
the manifold of classically degenerate states at finite temperatures. The formal procedure which we will be using here is based on the
derivation of the fluctuation-induced part of the free energy on top of the
mean field contribution, which then allows to determine the orientations of
the vector order parameter corresponding to the free energy minima.

In general, the calculation of the free energy of a quantum spin system is
complicated by the fact that the spin operators are non-canonical, which
limits the usefulness of renormalized perturbation theory, the usual tool in
dealing with quantum many-body systems.\cite{Negele1998,Fradkin2013} 
Here we explore a different approach: decoupling of the bilinear interaction operators
by way of Hubbard-Stratonovich (HS) transformations.\cite{hubb59,strat58} Our approach differs from a previous derivation  \cite{angelucci89,angelucci91,angelucci92} of a path integral representation of interacting quantum spin systems in terms of Hubbard-Stratonovich fields in that it applies to anisotropic spin systems and is not restricted to ultralow temperatures but allows to calculate the contribution of Gaussian fluctuations to the free energy at all temperatures in the ordered phase.  This
requires introducing space and time-dependent HS-fields, which are integrated
over. The resulting quantum trace over an exponential involving spin operators
(a kind of Zeeman interaction of spins with the HS-induced "magnetic field"),
may be done explicitly. The result is a representation of the partition
function in terms of an infinite power series in the interaction. The series
may be summed up explicitly in the case of an Ising-type interaction or in the
case of classical spins.\cite{sizyuk15,sizyuk16} In the general case of the bilinear
interaction of quantum spins, only the Gaussian fluctuation contribution to
the free energy of an ordered state may be derived in a compact form. Higher
order corrections are accessible, but involve increasingly complex expressions.

This paper is organized as follows: In Section II, the representation of the
partition function is introduced. In Section III we present the evaluation of
the free energy in the approximation of Gaussian fluctuations about the mean
field order parameter. We apply the derived formalism to the quantum
Heisenberg-compass spin model in section IV. Finally, we draw conclusions in
Section V.

\section{Representation of the partition function}

We consider a generic anisotropic quantum spin model on a lattice defined by
the Hamiltonian
\begin{equation}
{\mathcal{H}}=\frac{1}{2}\sum_{j,j^{\prime}}\,\sum_{\alpha\alpha^{\prime}%
}\,J_{j,j^{\prime}}^{\alpha\alpha^{\prime}}\,S_{j}^{\alpha}\,S_{j^{\prime}%
}^{\alpha^{\prime}}, \label{eq:ham}%
\end{equation}
where $j,j^{\prime}$ are combined indices, $j=(i,\mu)$, denoting the position
of the site by the position of the unit cell $i$\ and sublattice index $\mu$
inside the unit cell. $\alpha,\alpha^{\prime}=x,y,z$ label the three
components of the spin. In principle the $\mathbf{S}_{j}$ may be general spin
operators, but we will henceforth assume $S=1/2$, which is the most important
case. For the models with compasslike anisotropic and Heisenberg isotropic
interactions of spins, the interaction is diagonal in spin space,
$J_{j,j^{\prime}}^{\alpha\alpha^{\prime}}\propto\delta_{\alpha\alpha^{\prime}%
}$. However, since our consideration is also valid for the case when
$\alpha\neq\alpha^{\prime}$, in the following, we will keep both indices.

The partition function of the system is given by the trace over many-body spin
space of the Boltzmann operator
\begin{equation}
Z=\mathrm{Tr}\left[  \exp\left[  -\beta\frac{1}{2}\sum_{j,j^{\prime}}%
\,\sum_{\alpha\alpha^{\prime}}\,J_{j,j^{\prime}}^{\alpha\alpha^{\prime}%
}\,S_{j}^{\alpha}\,S_{j^{\prime}}^{\alpha^{\prime}}\right]  \right]  ,
\label{Z1}%
\end{equation}
where $\beta=1/k_{B}T$ is the inverse temperature (we will use units with
Boltzmann's constant $k_{B}=1$ and also Planck's constant $\hbar=1$).

It is useful to represent the Hamiltonian in the basis of the normalized
eigenfunctions, $\chi_{\mathbf{q},\nu;j,\alpha}$, of the spin exchange matrix
$J_{j,j^{\prime}}^{\alpha\alpha^{\prime}}$. Here we denote the lattice vectors
as $\mathbf{R}_{j}=\mathbf{R}_{i}+\mathbf{\tau}_{\mu}$, where $i=1,...,N$
specifies the unit cell and $\mathbf{\tau}_{\mu},$ $\mu=1,..,N_{s}$ denotes
the lattice vectors inside a unit cell. The eigenfunctions $\chi
_{\mathbf{q},\nu}$ and eigenvalues $\kappa_{\mathbf{q},\nu}$ are defined as%
\begin{equation}
\sum_{j^{\prime},\alpha^{\prime}}J_{j,j^{\prime}}^{\alpha\alpha^{\prime}}%
\chi_{\mathbf{q},\nu;j^{\prime},\alpha^{\prime}}=\kappa_{\mathbf{q},\nu}%
\chi_{\mathbf{q},\nu;j,\alpha}\label{eigenstates}%
\end{equation}
For spins on a periodic lattice these eigenstates are labeled by a wavevector
$\mathbf{q}$ from the first Brillouin zone (BZ) and index $\nu$, whose
dimensionality depends on the number $N_{s}$ of non-equivalent atoms in the
magnetic unit cell, $\nu=1,...3N_{s}$. Explicitly we have $\chi_{\mathbf{q}%
,\nu;j,\alpha}=N^{-1/2}\exp(i\mathbf{q\cdot R}_{i})u_{\mathbf{q}\nu}%
^{\mu\alpha}$\ , where the $u_{\mathbf{q}\nu}^{\mu\alpha}$ are the $3N_{s}%
$\ components (labeled $\mu\alpha$) of the normalized eigenvector labeled by
$\mathbf{q}\nu$ and $N$ is the number of lattice cells. Defining collective
spin operators ${\tilde{S}}_{\mathbf{q},\nu}$ we can express the Hamiltonian
as%
\begin{align}
{\mathcal{H}} &  =\sum_{\mathbf{q},\nu}\kappa_{\mathbf{q},\nu}{\tilde{S}%
}_{\mathbf{q},\nu}^{\ast}{\tilde{S}}_{\mathbf{q},\nu}\,,\label{H-momentum}\\
{\tilde{S}}_{\mathbf{q},\nu} &  =\sum_{j,\alpha}\chi_{\mathbf{q},\nu;j,\alpha
}^{\ast}S_{j}^{\alpha},\nonumber
\end{align}
where ${\bm\tilde{S}}_{\bm q}$ is a $3N_{s}-$component spin vector,
${\tilde{S}}_{\mathbf{q},\nu}^{\ast}={\tilde{S}}_{-\mathbf{q},\nu}$, and
$\kappa_{\mathbf{q},\nu}$ is the diagonal interaction matrix in momentum
space, which is obtained by Fourier transformation of the interaction matrix
$J_{j,j^{\prime}}^{\alpha\alpha^{\prime}}$ , and with the help of a unitary
transformation in the $3N_{s}$ dimensional space of spin components $\alpha$
and unit cell positions $\mu$. From the symmetry of the exchange interaction,
$J_{j,j^{\prime}}^{\alpha\alpha^{\prime}}=J_{j^{\prime},j}^{\alpha^{\prime
}\alpha}$\ it follows that $\kappa_{\mathbf{q,}\nu}=\kappa_{-\mathbf{q,}\nu}$.
For simplicity, in the following we will omit the tilde sign and put
$S_{\mathbf{q},\nu}\equiv{\tilde{S}}_{\mathbf{q},\nu}$.

In the eigenmode representation, the partition function of the system is given
by
\begin{equation}
Z=\mathrm{Tr}\left[  \exp\left[  -\beta\sum_{\mathbf{q},\nu}\kappa
_{\mathbf{q},\nu}{S}_{\mathbf{q},\nu}^{\ast}{S}_{\mathbf{q},\nu}\right]
\right]  , \label{Z2}%
\end{equation}

\subsection{Hubbard-Stratonovich transformation}

Our next step is to apply the Hubbard-Stratonovich transformation to each
normal component separately. This transformation is based on the mathematical
identitiy (defining ${x}=\operatorname{Re}\{{S}_{\mathbf{q},\nu}\},$
$y=\operatorname{Im}\{{S}_{\mathbf{q},\nu}\}$, and $a=\beta\kappa
_{\mathbf{q},\nu}$)
\begin{align}
\exp\left[  -a(x^{2}+y^{2})\right]   &  =\frac{1}{\pi|a|}\int dudv
\label{HS1}\\
&
\genfrac{\{}{.}{0pt}{}{\exp\left[  -\frac{u^{2}+v^{2}}{|a|}+2(xu+yv)\right]
\text{, \ \ \ \ \ \ }a<0,}{\exp\left[  -\frac{u^{2}+v^{2}}{|a|}%
+2i(-xu+yv)\right]  \text{, \ \ }a>0}%
,\nonumber
\end{align}
In the second equality we made use of the freedom to choose the imaginary
prefactor to be $+i$ or $-i$ . Here the auxiliary variables $u,v$ are
components of the Hubbard-Stratonovich field $\varphi_{\mathbf{q,}\nu}$, which
we choose to be $u=\operatorname{Re}\varphi_{\mathbf{q,}\nu}$ and
$v=\operatorname{Im}\varphi_{\mathbf{q,}\nu}$ in the first equation (valid if
$a<0$) and $v=\operatorname{Re}\varphi_{\mathbf{q,}\nu}$ and
$u=\operatorname{Im}\varphi_{\mathbf{q,}\nu}$ in the second equation (valid if
$a>0$). We may combine both cases by defining a function $s(\kappa
_{\mathbf{q,}\nu})=1$ if $\kappa_{\mathbf{q,}\nu}<0$ and $s(\kappa
_{\mathbf{q,}\nu})=-i$ if $\kappa_{\mathbf{q,}\nu}>0$\ . Then in the first
equation $(xu+yv)=\operatorname{Re}\{s(\kappa_{\mathbf{q,}\nu}){S}%
_{\mathbf{q},\nu}\varphi_{\mathbf{q,}\nu}^{\ast}\},$ whereas in the second
equation $i(-xu+yv)=\operatorname{Im}\{{S}_{\mathbf{q},\nu}\varphi
_{\mathbf{q,}\nu}^{\ast}\},$ which may also be expressed as $\operatorname{Re}%
\{s(\kappa_{\mathbf{q,}\nu}){S}_{\mathbf{q},\nu}\varphi_{\mathbf{q,}\nu}%
^{\ast}\}$ . In both cases, whether $s(\kappa_{\mathbf{q},\nu})$ is real or
imaginary valued, the term in the exponent linear in ${S}_{\mathbf{q},\nu}$ is
real valued. In Eq.(\ref{HS1}), $x+iy$ may be a number or an operator.

The application of the Hubbard-Stratonovich transformation to (\ref{Z2})
requires the normal components of the spin operators to commute with each
other, which is true for classical spins. Then using the Hubbard-Stratonovich
transformation one may express the Boltzmann weight operator of each normal
mode in terms of classical fields $\varphi_{\mathbf{q,}\nu}$ and represent the
interaction operator as a Zeeman energy operator of spins in a spatially
varying magnetic field \cite{sizyuk15}. Note that in \cite{sizyuk15} we used a
slightly different convention for the definition of variables $u,v$ above.

In order to make use of the Hubbard-Stratonovich transformation for the
computation of the partition function (\ref{Z2}) of a quantum spin system, for
which the $S_{\mathbf{q},\nu}$ do not commute, we need to represent the
Boltzmann operator as an evolution operator in imaginary time and apply the
Suzuki-Trotter discretization \cite{trott59}, allowing to write down the
partition function in terms of products over time slices.\cite{Negele1998} 
Explicitly, we have
\begin{align}
Z  &  =\mathrm{Tr}\left[  \exp\left(  -\beta\mathcal{H}\right)  \right]
\label{Z3}\\
&  =\mathrm{Tr}\left[  T_{\tau}\exp[-\epsilon\sum_{n=1}^{M}\mathcal{H}(\tau_{n})] \right]  ,\nonumber
\end{align}
where $T_{\tau}$ is the imaginary time ordering operator, and we sliced the
imaginary time interval $[0,\beta]$ into $M=\beta/\epsilon$ infinitesimal
intervals each of length $\epsilon$, and $\tau_{n}=n\epsilon,$ $n=1,2,...M$.
Since $\epsilon$ is small, and will be taken to zero at the end,\ we may now
expand each exponent in Eq. (\ref{Z3}) as
\begin{equation}
\exp[-\epsilon\mathcal{H}(\tau_{n})]=1-\epsilon\mathcal{H}(\tau_{n}%
)+O(\epsilon^{2}),
\end{equation}
and approximate
\begin{equation}
T_{\tau}\exp[-\epsilon\sum_{n=1}^{M}\mathcal{H}(\tau_{n})]=T_{\tau}\prod
_{n=1}^{M}\exp[-\epsilon\mathcal{H}(\tau_{n})],
\end{equation}
where
\begin{align}
&  \exp\left[  -\epsilon\mathcal{H}(\tau_{n})\right]  =\exp\left[
-\epsilon\sum_{\mathbf{q},\nu}\kappa_{\mathbf{q},\nu}S_{\mathbf{q},\nu}^{\ast
}(\tau_{n})S_{\mathbf{q},\nu}(\tau_{n})\right] \nonumber\\
&  =\prod_{\mathbf{q},\nu}\exp[-\epsilon\kappa_{\mathbf{q},\nu}S_{\mathbf{q}%
,\nu}^{\ast}(\tau_{n})S_{\mathbf{q},\nu}(\tau_{n})]+O(\epsilon^{2}).
\end{align}
Now, at each instant $\tau_{n}$ we may perform the Hubbard-Stratonovich
transformation separately and express the Boltzmann weight operator of each
normal mode in terms of normal field amplitudes $\varphi_{\mathbf{q,}\nu}%
(\tau_{n})$ as%
\begin{align}
&  \exp\left[  -\epsilon\kappa_{\mathbf{q,}\nu}S_{\mathbf{q,}\nu}^{\ast}%
(\tau_{n})S_{\mathbf{q,}\nu}(\tau_{n})\right]  =\\
&  C_{\mathbf{q,}\nu}^{-1}\int d\varphi_{\mathbf{q,}\nu}^{\ast}d\varphi
_{\mathbf{q,}\nu}\exp\Bigl[-\epsilon{\Bigl\{}|\kappa_{\mathbf{q,}\nu}%
|^{-1}\varphi_{\mathbf{q,}\nu}^{\ast}(\tau_{n})\varphi_{\mathbf{q,}\nu}%
(\tau_{n})\nonumber\\
&  -2\operatorname{Re}\{s(\kappa_{\mathbf{q,}\nu})S_{\mathbf{q,}\nu}^{\ast
}(\tau_{n})\varphi_{\mathbf{q,}\nu}(\tau_{n})\}{\Bigr\}}\Bigr].\nonumber
\end{align}

The partition function (\ref{Z3}) may therefore be expressed
as\begin{widetext}
\begin{eqnarray}\label{Z4}
&&
Z =\frac{1}{C}{\rm Tr}
\left[T_{\tau}
\int\left[d\varphi\right]
\prod_{n=1}^{M}\exp
\left[
-\epsilon
\left \{
\sum_{\mathbf{q},\nu}
\left(
|\kappa_{\mathbf{q},\nu}|^{-1}
\varphi_{\mathbf{q},\nu}^{\ast}(\tau_{n})
\varphi_{\mathbf{q},\nu}(\tau_{n})
-2\operatorname{Re}\{s(\kappa_{\mathbf{q,}\nu})S_{\mathbf{q,}\nu}^{\ast
}(\tau_{n})\varphi_{\mathbf{q,}\nu}(\tau_{n})\}
\right)
\right \}
\right]
\right]
\nonumber\\
&& =\frac{1}{C}
\int\left[d\varphi\right]\prod_{n=1}^{M}
\exp\left[
-\epsilon
\sum_{\mathbf{q},\nu}
|\kappa_{\mathbf{q},\nu}|^{-1}
\varphi_{\mathbf{q},\nu}^{\ast}(\tau_{n})
\varphi_{\mathbf{q},\nu}(\tau_{n})\right]
{\rm Tr}
\left[
T_{\tau}\prod_{j,\alpha}
\exp\left[
-\epsilon B^{\alpha}_{j,{\mathrm eff}}
(\tau_{n})S_{j}^{\alpha}(\tau_{n})
\right]
\right],
\end{eqnarray}
\end{widetext}where the spatially and "temporally" varying local magnetic
field $B_{j,\mathrm{eff}}^{\alpha}(\tau_{n})$ is defined by
\begin{align}
B_{j,\mathrm{eff}}^{\alpha}(\tau_{n})  &  =-2\varphi_{j}^{\alpha}(\tau
_{n})\label{field-fourier}\\
\varphi_{j}^{\alpha}(\tau_{n})  &  =\sum_{\mathbf{q},\nu=1}^{3N_{s}%
}\operatorname{Re}\{s(\kappa_{\mathbf{q,}\nu})\chi_{\mathbf{q},\nu;j,\alpha
}^{\ast}\varphi_{\mathbf{q,}\nu}(\tau_{n})\}.\nonumber
\end{align}
\ In Eq. (\ref{Z4}), we also defined
the integration volume element as
\[
\lbrack d\varphi]=\Pi_{\mathbf{q,}\nu,n}{d\varphi_{\mathbf{q,}\nu}^{\ast}%
}(\tau_{n})d\varphi_{\mathbf{q,}\nu}(\tau_{n})
\]
and the normalization factor $C$ as
\[
C=\int[d\varphi]\prod_{n=1}^{M}\exp\left[  -\epsilon\sum_{\mathbf{q},\nu
}|\kappa_{\mathbf{q},\nu}|^{-1}\varphi_{\mathbf{q},\nu}^{\ast}(\tau
_{n})\varphi_{\mathbf{q},\nu}(\tau_{n})\right]
\]

\subsection{Trace over quantum spin states}

We may now perform the trace over the quantum spin states. As spin operators
on different sites commute, the trace may be split up into a product of traces
$\mathrm{Tr}_{j}$ over single spin Hilbert space
\begin{align}
&  Z=\frac{1}{C^{\prime}}\int[d\varphi]\label{Z5}\\
&  \exp\left[  -\int_{0}^{\beta}d\tau\sum_{\mathbf{q},\nu}|\kappa
_{\mathbf{q,}\nu}|^{-1}\varphi_{\mathbf{q,}\nu}^{\ast}(\tau)\varphi
_{\mathbf{q,}\nu}(\tau)\right]  \prod_{j}\Theta_{j},\nonumber
\end{align}
where we defined
\begin{equation}
\Theta_{j}=\frac{1}{2}\mathrm{Tr}_{j}{\Bigl \{}T_{\tau}\prod_{n}\exp\left[
2\epsilon\sum_{\alpha=x,y,z}\varphi_{j}^{\alpha}(\tau_{n})S_{j}^{\alpha
}\right]  {\Bigr \}} \label{Thetaj}~.%
\end{equation}
The factors $\frac{1}{2}$ in front of the trace are compensated by
corresponding factors in the normalization, leading to $C^{\prime}=C/2^{N_{t}%
}$, where $N_{t}$ is the total number of spins.

We note that the factors under the trace, $\exp\left[  2\epsilon
\boldsymbol{\varphi}_{j}(\tau_{n})\cdot\mathbf{S}_{j}\right]  $, may be
cyclically permuted. This suggests that the terms at $n=1$ and at $n=M$ should
be equal, in order to avoid an unphysical discontinuity when passing from
$n=1$ to $n=M$. In other words, we assume periodic boundary conditions,
$\varphi_{\mathbf{q},\nu}(\tau_{n})=\varphi_{\mathbf{q},\nu}(\tau_{n}+\beta)$. This implies that the Fourier frequencies $\omega_{n}$ of $\varphi
_{\mathbf{q},\nu}(\tau)$ are bosonic Matsubara frequencies. \ 

The spin trace may be performed by inserting suitable representations of the
unit operator in single spin space at each time step%
\begin{equation}
\hat{\mathbf{1}}=\sum_{s=\pm1}|s;\mathbf{e}_{n}\rangle\langle s;\mathbf{e}_{n}|,
\end{equation}
where $\mathbf{e}_{n}\equiv \mathbf{e}(\tau_n)= \boldsymbol{\varphi}_{j}(\tau_{n})/|\boldsymbol{\varphi}%
_{j}(\tau_{n})|$ denotes the spin quantization axis at time $\tau_{n}$ and
$ |s;\mathbf{e}_{n}\rangle\equiv|s_{n}\rangle $, $s=\pm1$ are  corresponding two-component
spin eigenvectors, obeying $(\mathbf{S\cdot e}_{n})|s_{n}\rangle=\frac{s}{2}|s_{n}\rangle$. The assumed
periodic boundary condition implies that $\mathbf{e}_{N}=\mathbf{e}_{1}$ such
that $\mathbf{e(}\beta)=\mathbf{e(}0)$\ and $\mathbf{e(}\tau)$ is a periodic
function of period $\beta$. 

 Now, inserting  the unit operator
expressed in the basis of eigenstates of each factor ${e^{\epsilon
\boldsymbol{\varphi}_{n}\cdot\boldsymbol{\sigma}}}$ to the left and right of that
factor in the Eq. \ref{Thetaj}, we may express $\Theta_{j}$ as
\begin{align}
\Theta &  =\frac{1}{2}\mathrm{Tr}{\Bigl \{e^{\epsilon\boldsymbol{\varphi}%
_{N}\cdot\boldsymbol{\sigma}}.......e^{\epsilon\boldsymbol{\varphi}_{2}\cdot
\boldsymbol{\sigma}}e^{\epsilon\boldsymbol{\varphi}_{1}\cdot\boldsymbol{\sigma}}%
\Bigr \}}\label{traceB}\nonumber\\ 
&  =\frac{1}{2}\sum_{s_{1},s_{1}^{\prime},s_{2},..}\langle s_{1}%
|e^{\epsilon\boldsymbol{\varphi}_{N}\cdot\boldsymbol{\sigma}}|s_{N}^{\prime}%
\rangle\langle s_{N}^{\prime}{|...}\\\nonumber
&  ..{.|}s_{2}^{\prime}\rangle\langle s_{2}^{\prime}|e^{\epsilon
\boldsymbol{\varphi}_{2}\cdot\boldsymbol{\sigma}}|s_{2}\rangle\langle s_{2}%
|s_{1}^{\prime}\rangle\langle s_{1}^{\prime}|e^{\epsilon\boldsymbol{\varphi}%
_{1}\cdot\boldsymbol{\sigma}}|s_{1}\rangle,
\end{align}
where for brevity we    suppressed the site index $j$ and
defined $\boldsymbol{\varphi}_{n}=\boldsymbol{\varphi}(\tau_{n})$,  where  $\boldsymbol{\varphi=(}\varphi^{x},\varphi
^{y},\varphi^{z}\mathbf{)}$ is
 the vector in spin space. 
 
 Next, we need to
compute the matrix elements $\langle s_{n}^{\prime}|e^{\epsilon\boldsymbol{\varphi
}_{1}\cdot\boldsymbol{\sigma}}|s_{n}\rangle$ and the inner products $\langle
s_{n+1}|s_{n}^{\prime}\rangle$. The former are diagonal by construction:%
\begin{equation}
\langle s_{n}^{\prime}|e^{\epsilon\boldsymbol{\varphi}_{n}\cdot\boldsymbol{\sigma}%
}|s_{n}\rangle=\delta_{s_{n}s_{n}^{\prime}}\exp[\epsilon\varphi_{n}%
\sigma_{s_{n}s_{n}}^{z}],
\end{equation}
where $\varphi_{n}=|\mathbf{\varphi(}\tau_{n})|=\sqrt{(\varphi^{x})^{2}%
+(\varphi^{y})^{2}+(\varphi^{z})^{2}}$. 
The inner product is given by%
\begin{align}
&\langle s_{n+1}|s_{n}^{\prime}\rangle  = \langle s_{n+1}, \mathbf{e}_{n+1}|s_{n}^{\prime}, \mathbf{e}_{n}\rangle\nonumber\\
& =\langle s_{n+1}|\exp[-i(\mathbf{e}%
_{n+1}\times\mathbf{e}_{n})\cdot \boldsymbol{\sigma}]|s_{n}^{\prime
};\mathbf{e}_{n+1}\rangle\\\nonumber
&  =\delta_{s_{n+1}s_{n}^{\prime}}-i(\mathbf{e}_{n+1}\times\mathbf{e}%
_{n})\cdot\boldsymbol{\sigma}_{s_{n+1}s_{n}^{\prime}}+O(\epsilon^{2}).
\end{align}

 Next we define the infinitesimal angle of rotation of the quantization axis
$\mathbf{e}_{n+1}$ into $\mathbf{e}_{n}$ as $\epsilon\,
\mathbf{\Omega}_{n+1,n}=(\mathbf{e}_{n}\times\mathbf{e}_{n+1})$ and express
$\Theta$ as \
\begin{align}\nonumber
\Theta &  =\frac{1}{2}Tr\{e^{\epsilon\varphi_{N}\sigma^{z}}e^{i\epsilon
\mathbf{\Omega}_{1,N-1}\cdot\boldsymbol{\sigma}}e^{\epsilon\varphi_{N-1}\sigma
^{z}}...\\
&  ...e^{i\epsilon\mathbf{\Omega}_{3,2}\cdot\boldsymbol{\sigma}}e^{\epsilon
\varphi_{2}\sigma^{z}}e^{i\epsilon\mathbf{\Omega}_{2,1}\cdot\boldsymbol{\sigma}%
}e^{\epsilon\varphi_{1}\sigma^{z}}\},\label{Theta- intermediate}
\end{align}
where we denote the quantization axis at time $\tau=0$ as $\mathbf{e}%
_{1}=\mathbf{e}(0)=\widehat{z}$. In the continuum approximation, we have 
\begin{equation}
\mathbf{e}_{n+1}=\mathbf{e(}\tau
_{n+1})=\mathbf{e(}\tau_{n})+\epsilon\mathbf{\dot{e}(}\tau_{n}%
)+O(\epsilon^{2}), 
\end{equation}
where $\mathbf{\dot{e}(}\tau)=\partial\mathbf{e}%
/\partial\tau$. 

 The factors in the product on the r.h.s. of the Eq.(\ref{Theta- intermediate}) commute under the time ordering operator,  so we may now express $\Theta$ as%
\begin{equation}
\Theta=\frac{1}{2}\mathrm{Tr}{\Bigl \{}T_{\tau}\exp\left[  \int_{0}^{\beta
}d\tau\{i\mathbf{\Omega(}\tau)
\mathbf{+}%
\varphi\mathbf{(}\tau\mathbf{)e(}0\mathbf{)\}}\cdot\boldsymbol{\sigma}\right]
{\Bigr \}},%
\end{equation}
 where $\mathbf{\Omega(}\tau_{n})=\mathbf{\Omega}%
_{n+1,n}=\mathbf{e(}\tau_{n})\times\mathbf{\dot{e}(}\tau_{n})$. The term $ i\Omega (\tau)$ may be identified with the well-known Berry phase term.

Although a  general evaluation of this expression involves only a trace
over a single spin,  it appears  to be difficult. We therefore consider  the following approximations.
The field $\mathbf{\Omega}$ is by definition a fluctuation field, i.e. it
vanishes on the mean field level. It therefore makes sense to expand $\Theta$
in terms of $\mathbf{\Omega}$. In zeroth order we have
\begin{align}
\Theta^{(0)}  &  =\frac{1}{2}Tr\{e^{\epsilon\varphi_{N}\sigma^{z}}%
e^{\epsilon\varphi_{N-1}\sigma^{z}}.....e^{\epsilon\varphi_{2}\sigma^{z}%
}e^{\epsilon\varphi_{1}\sigma^{z}}\}\label{theta_zero}\\
&  =\frac{1}{2}Tr\{e^{\int d\tau\varphi(\tau)\sigma^{z}}\}=\cosh\beta
\varphi_0,\nonumber
\end{align}
where the time average $\varphi_0$ is defined as%
\begin{equation}
\varphi=\beta^{-1}\int_{0}^{\beta}d\tau\varphi(\tau). \label{phi_static}%
\end{equation}
We note for later that $\Theta^{(0)}$ contains contributions from both longitudinal
and transverse fluctuations about the mean field configuration.\ 

In first order in $\mathbf{\Omega}$ we find%
\begin{align}
\Theta^{(1)} &  =\frac{1}{2}i\epsilon\sum_{n_{0}=1}^{N-1}Tr\{e^{
\epsilon\sigma^{z}\sum_{n_{2}=n_{0}+1}^{N}\varphi_{n_{2}}}[\mathbf{\Omega
}_{n_{0}+1,n_{0}}\cdot\boldsymbol{\sigma}]\nonumber\\
&  \times e^{\epsilon\sigma^{z}\sum_{n_{1}=1}^{n_{0}}\varphi_{n_{1}}}\}.
\end{align}
Higher order contributions in $\Omega$  may be derived but they lead to increasingly complicated expressions.
In the continuum approximation we may express $\Theta^{(1)}$ as\
\begin{equation}
\Theta^{(1)}=\frac{i}{2}\int d\tau_0 Tr\{e^{\sigma^{z}\Phi(\beta
,\tau_0)}[\mathbf{\Omega(}\tau_0\mathbf{)}\cdot\boldsymbol{\sigma}]e^{\sigma^{z}%
\Phi(\tau_0,0)}\},
\end{equation}
where $\Phi(\beta,\tau_{0})=\int_{\tau_{0}}^{\beta}d\tau\varphi(\tau)$
has been defined. We now recall that $\sigma^{z}=\mathbf{e(}0)\cdot
\boldsymbol{\sigma}$ and use $e^{\sigma^{z}\Phi}=\cosh\Phi+\sigma^{z}\sinh\Phi$ to get%
\begin{align}
\Theta^{(1)}  &  =i\int d\tau_{0}[\cosh\Phi(\beta,\tau_{0})\sinh\Phi(\tau
_{0},0)\nonumber\\
&  +\sinh\Phi(\beta,\tau_{0})\cosh\Phi(\tau_{0},0)][\mathbf{\Omega(}\tau
_{0}\mathbf{)}\cdot\mathbf{e(}0)], 
\end{align}
 where we also used $Tr\{\boldsymbol{\sigma}\}=0$, $Tr\{\sigma^{z}\sigma^{\lambda
}\mathbf{\}=}2\delta_{\lambda,z}=Tr\{\sigma^{\lambda}\sigma^{z}\mathbf{\}}$ and 
$Tr\{\sigma^{z}\sigma^{\lambda}\sigma^{z}\}=0$.
Using another identity, $\sinh x\cosh y+\cosh x\sinh y=\sinh(x+y)$, and the periodic boundary condition relations  $\Phi(\beta,\tau
_{0})+\Phi(\tau_{0},0)=\Phi(\beta,0)=\beta\varphi_{0}$  we get
\begin{equation}
\Theta^{(1)}=i[\mathbf{\Omega}_{0}\cdot\mathbf{e(}0)]\sinh(\beta\varphi_0),
\end{equation}
where 
$\mathbf{\Omega}_{0}\equiv\int d\tau_{0}\mathbf{\Omega(}\tau_{0}\mathbf{)}$.\

The partition function Eq. (\ref{Z5}), which includes   the first order  correction in $\mathbf{\Omega}_{0}$, is then given by%
\begin{equation}
Z   =\frac{1}{C^{\prime}}\int[d\varphi]\exp[-\beta({\mathcal{S}}_{\kappa
}+{\mathcal{S}}_{loc}+{\mathcal S}_{0})],
\end{equation}
where the interaction part of the action is given by
\begin{equation}
{\mathcal{S}}_{\kappa}   =\beta^{-1}\int_{0}^{\beta}d\tau\sum_{\mathbf{q}%
,\nu}|\kappa_{\mathbf{q,}\nu}|^{-1}\varphi_{\mathbf{q,}\nu}^{\ast}%
(\tau)\varphi_{\mathbf{q,}\nu}(\tau)
\end{equation}
 and the local part of the action $ {\mathcal{S}}_{loc}  ={\mathcal{S}}_{loc}^{stat}+{\mathcal{S}}_{loc}
^{dyn}$ has both
 static and dynamic contributions: 
\begin{align}
S_{loc}^{stat}  &  =-\beta^{-1}\sum_{j}\ln\cosh(\beta\varphi_{j,0})\\
S_{loc}^{dyn}  &  =-i\beta^{-1}\sum_{j}\tanh(\beta\varphi_{j,0}%
)[\mathbf{\Omega}_{j,0}\cdot\mathbf{e}_{j}\mathbf{(}0)],
\end{align}
 and ${\mathcal S}_{0}=\beta^{-1}\ln C^{\prime}$.

Our results in Eqs.(29-32) agree with those of Ref.\cite{angelucci89} except for an additional term involving the product of two time derivatives of the transverse field components at equal times. Such a term arises from expansion of Eq.(22) in second order in $\Omega$, if the time arguments are kept equal. The fluctuations we will be interested in (e.g. spin waves) are long-range correlated in time such that it does not make sense to single out only the equal time products of $\Omega$. We also note that  Angelucci and Jug\cite{angelucci89}  did not pay attention to the fact that the Hubbard-Stratonovich transformation changes its character if the eigenvalues $\kappa_{\mathbf{q,}\nu}$ of the interaction kernel change sign.

\section{Mean field solution}

In this section,  we consider  the simplest case and compute the mean field free energy for the range of
parameters of a model (\ref{eq:ham}), for which the mean-field solution is a
collinear magnetic state. In this case we can write $\boldsymbol{\varphi}_{j,\mu}(\tau)=\varphi
_{\small{\rm  MF}}\hat{\mathbf{m}}$, where $\hat{\mathbf{m}}$ is a normalized $3$-component
vector pointing in the direction of the spontaneous magnetization, which is the same for all 
sublattice sites $\mu$. 
The trace in spin space is
obtained as
\[
\Theta_{j,\mu}^{\mathrm{MF}}=\frac{1}{2}\mathrm{Tr}_{j}\{e^{-\beta{\varphi
}_{\mathrm{MF}}[\boldsymbol{\sigma}\cdot\mathbf{m}]}\}=\cosh(\beta
\varphi_{\mathrm{MF}}),
\]
where the mean field expression for the fields $\varphi_{\mathbf{q},\nu}(\tau)$ is
given by $\varphi_{\mathbf{q},\nu}^{MF}=(NN_{s})^{1/2}\delta_{\mathbf{q}%
,0}\varphi_{\mathrm{MF}}m_{\mathbf{q},\nu}$, where  $N$ is the total number of unit cells and $N_s$  is  three times the number of the sublattices in the unit cell. 
The normalized
unit vector $m_{\mathbf{q},\nu}=N_{s}^{-1/2}\sum_{\mu,\alpha}m^{\alpha
}u_{\mathbf{q},\nu}^{\mu\alpha}$  is expressed in terms of the eigenvectors $u_{\mathbf{q}%
,\nu}^{\mu\alpha}$.
The mean-field partition function (\ref{Z5}) can be easily evaluated and equals
\begin{equation}
Z^{\mathrm{MF}}=\frac{1}{C^{\prime}}\exp\left[  -NN_{s}\left\{  \beta
|\kappa_{0}|^{-1}\varphi_{\mathrm{MF}}^{2}-\ln\left(  \cosh(\beta
\varphi_{\mathrm{MF}})\right)  \right\}  \right]  \label{Z_MF},%
\end{equation}
 provided $\kappa_{\mathbf{q}=0,\nu}=\kappa_{0}$ is  independent of $\nu$.\ Here
 $\varphi_{\mathrm{MF}}$ is the value minimizing the free energy
$F^{\mathrm{MF}}=-\beta^{-1}\ln Z^{\mathrm{MF}}$ and is\ given by the solution
of the transcendental equation%
\begin{equation}
2|\kappa_{0}|^{-1}\varphi_{\mathrm{MF}}=\tanh(\beta\varphi_{\mathrm{MF}}).
\end{equation}
The full partition function is given by
\[
Z=\frac{Z^{\mathrm{MF}}}{C^{\prime}}\int[d\delta\varphi]\exp\left[
-\beta\delta{\mathcal{S}}\right]  ,
\]
where $\delta{\mathcal{S=}}\delta{\mathcal{S}}^{stat}{\mathcal{+}}%
\delta{\mathcal{S}}^{dyn}$ is the fluctuational part of the action.

Despite the anisotropic form of the interactions  in the Hamiltonian(\ref{eq:ham}), the
mean-field solution is highly degenerate with respect to the orientation of
the spontaneous magnetization vector $\hat{\mathbf{m}}$. It is therefore of
interest to calculate the corrections to the mean-field solution capturing the
anisotropy of the free energy with respect to the order parameter orientation.

\section{ Evaluation of the free energy in the Gaussian approximation}

The first systematic free energy correction is that from \textit{Gaussian
fluctuations} about the mean-field  solution obtained by expanding the free energy, or
equivalently the action, to lowest order in the fluctuation field
$\delta{\varphi}_{\mathbf{q},\nu}(\tau)={\varphi}_{\mathbf{q},\nu}(\tau
)-{\varphi}_{\mathbf{q},\nu}^{\mathrm{MF}}$. Introducing the time Fourier
transform
\[
{\varphi}_{\mathbf{q},\nu}(\tau)=\sum_{\omega_{n}}{\varphi}_{\mathbf{q},\nu
,\omega_{n}}\exp\left[  i\omega_{n}\tau\right],
\]
where $\omega_{n}=2\pi n\beta^{-1}$ are bosonic Matsubara frequencies,  we immediately get the following bilinear
form of  the Gaussian fluctuation part of the action:
\begin{equation}
\delta{\mathcal{S}}\{\delta{\varphi}_{\mathbf{q,}\nu}\}   =\beta^{-1}%
\sum_{\mathbf{q,}\nu,\nu^{\prime}}\sum_{\omega_{n}}A_{\mathbf{q,}\omega
_{n};\nu\nu^{\prime}}\delta{{\varphi}}_{\mathbf{q,-}\omega_{n},\nu}^{(+)\ast
}\delta{{\varphi}}_{\mathbf{q,}\omega_{n},\nu^{\prime}}^{(+)},\label{fluct-action}%
\end{equation}
where
\begin{equation}
\delta{\varphi}_{\mathbf{q,}\omega_{n},\nu}^{(+)}   =\frac{1}{2}%
[s(\kappa_{\mathbf{q},\nu})\delta\varphi_{\mathbf{q,}\omega_{n},\nu}+
s^{\ast}(\kappa_{\mathbf{q},\nu})\delta\varphi_{-\mathbf{q,}\omega_{n},\nu
}^{\ast}].\label{varphi+}
\end{equation}
Here the fluctuation matrix elements $A_{\mathbf{q,}\omega_{n};\nu
\nu^{\prime}}$\ describe the weight of the Gaussian fluctuations of wavevector
$\mathbf{q}$, frequency $\omega_{n}$ and polarization $\nu$. 

 Here a comment is in order. The fields
$\delta\varphi$ do not obey the relation $\delta\varphi_{\mathbf{q,}\nu}%
^{\ast}(\tau_{1})=\delta\varphi_{-\mathbf{q,}\nu}(\tau_{1})$, i.e. their
spatial Fourier transforms are not real-valued, which leads to the combination
of fluctuation amplitudes
at momenta $\mathbf{q}$ and $-\mathbf{q}$, weighted by the phase factors
$s(\kappa_{\mathbf{q},\nu})$. Only symmetric combination of the fields,  gives contribution to the free energy  because the antisymmetric combination with  
$\delta \varphi_{\mathbf{q,}\omega_{n},\nu}^{(-)}=\frac{1}{2}%
[s(\kappa_{\mathbf{q},\nu})\delta\varphi_{\mathbf{q,}\omega_{n},\nu}-
s^{\ast}(\kappa_{\mathbf{q},\nu})\delta\varphi_{-\mathbf{q,}\omega_{n},\nu
}^{\ast}]$ drops out.

\subsection{Static fluctuations}

We start by considering the contribution of static Gaussian fluctuations to
the free energy. The
expansion of ${\mathcal{S}}_{loc}^{\mathrm{stat}}$ in terms of fluctuation
amplitudes up to second order is given by\
\begin{align}
\delta{\mathcal{S}}_{loc}^{\mathrm{stat}} &  =-\beta^{-1}\delta\{\sum_{j}%
\ln\cosh(\beta\lbrack(\varphi_{\mathrm{MF}}\hat{\mathbf{m}}+
\delta\boldsymbol{\varphi}_{i\mu,0})^{2}]^{1/2})\\
&  =-\frac{1}{2}\sum_{j}\{\beta_{c}\delta\boldsymbol{\varphi}_{i\mu,0}^{2}%
+\beta_{m}(\hat{\mathbf{m}}\cdot\delta\boldsymbol{\varphi}_{i\mu,0})^{2}\},\nonumber
\end{align}
where  $\beta_{m}=(1-t^{2})\beta-\beta_{c}$ and  $\beta_{c}=\frac{1}{T_{c}}$.
 Here
$t=\tanh(\beta\varphi_{\mathrm{MF}})$ denotes the dimensionless
measure of magnetization, which is zero at $T_c$ and rises
monotonically upon cooling to the saturation magnetization ($t=1$) at $T=0$.
 The components of $\delta
\boldsymbol{\varphi}_{i\mu,0}$ may be expressed in terms of the momentum space
fluctuation amplitudes $\delta\varphi_{\mathbf{q,}0,\nu}=\varphi
_{\mathbf{q,}0,\nu}-\varphi_{\mathbf{q},\nu}^{MF}$ as
\begin{equation}
\delta\varphi_{i\mu,0}^{\alpha}=\sum_{\mathbf{q},\nu}\operatorname{Re}%
\{s(\kappa_{\mathbf{q,}\nu})\chi_{\mathbf{q},\nu;j\alpha}^{\ast}\delta
\varphi_{\mathbf{q,}\nu,0}\}.
\end{equation}

Now we  can rewrite $ \delta{\mathcal{S}}_{loc}^{\mathrm{stat}} $ in the same   form as in  Eq.(\ref{fluct-action}): 
\begin{equation}
\delta{\mathcal{S}}^{\mathrm{stat}}=\beta^{-1}\sum_{\mathbf{q};\nu,\nu
^{\prime}}A_{\mathbf{q,}\nu\nu^{\prime}}^{\mathrm{stat}}\delta\varphi
_{\mathbf{q},0,\nu}^{(+)\mathrm{\ast}}\delta\varphi_{\mathbf{q},0,\nu^{\prime
}}^{(+)},
\end{equation}
 and the matrix $A_{\mathbf{q,}\nu\nu^{\prime}}^{\mathrm{stat}}$ is found to be
\begin{equation}
A_{\mathbf{q,}\nu\nu^{\prime}}^{\mathrm{stat}}=\beta\lbrack(|\kappa
_{\mathbf{q,}\nu}|^{-1}-\frac{\beta_{c}}{2})\delta_{\nu\nu^{\prime}}%
-\frac{\beta_{m}}{2}m_{\mathbf{q}\nu}m_{\mathbf{q}\nu^{\prime}}]\label{A_stat}.%
\end{equation}

Now the integration over the fluctuation amplitudes may be performed, with the
result
\begin{align}
Z^{\mathrm{stat}}=&\frac{\,Z^{\mathrm{MF}}}{C^{\prime}}\int[d\varphi
]\exp\left[  -\beta\delta{\mathcal{S}}^{\mathrm{stat}}\right]\nonumber\\
&=Z_{\mathrm{MF}}\exp\left[  -\beta\delta{\mathcal{F}%
}^{\mathrm{stat}}\right]  ,
\end{align}
which gives the free energy contribution to be equal to
\begin{equation}
\delta{\mathcal{F}}^{\mathrm{stat}}=\frac{1}{2}\beta^{-1}\sum_{\mathbf{q}}\ln
\det\{A_{\mathbf{q,}\nu\nu^{\prime}}^{\mathrm{stat}}\}.
\end{equation}

\subsection{Dynamic fluctuations}

We now turn to the dynamic fluctuations, obtained by expanding ${\mathcal{S}%
}_{loc}^{\mathrm{dyn}}$ to quadratic order in the finite frequency Fourier components $\delta
\boldsymbol{\varphi}_{j,\omega_{n}}$ of  the time-dependent fluctuation
fields.  
First we note that $\mathbf{\Omega}_{0}=\int d\tau_{0}\mathbf{\Omega(}\tau_{0}\mathbf{)}$ may
be expressed in terms of the transverse fluctuation amplitudes $\delta
\varphi_{j}^{tr,\alpha}(\tau)=\sum_{\alpha^{\prime}}P_{\alpha\alpha^{\prime}%
}\delta\varphi_{j}^{\alpha^{\prime}}(\tau)$, where $P_{\alpha\alpha^{\prime}%
}=\delta_{\alpha\alpha^{\prime}}-m_{\alpha}m_{\alpha^{\prime}}$, as%
\begin{align}\label{Omega0}
\mathbf{\Omega}_{j,0} &  =\int_{0}^{\beta}d\tau\delta\mathbf{e_j(}\tau
\mathbf{)\times}\delta\mathbf{\dot{e}_j(}\tau)\nonumber\\
&  =\frac{1}{\varphi_{MF}^{2}}\int_{0}^{\beta}d\tau\delta\boldsymbol{\varphi}%
_{j}^{\mathrm{tr}}(\tau)\mathbf{\times}\frac{\partial}{\partial\tau}%
\delta\boldsymbol{\varphi}_{j}^{\mathrm{tr}}\mathbf{(}\tau)\\
&  \mathbf{=}\frac{\beta_{c}^{2}\beta}{t^{2}}\sum_{\omega_{n}}i\omega
_{n}\delta\boldsymbol{\varphi}_{j,-\omega_{n}}^{\mathrm{tr}}\times\delta
\boldsymbol{\varphi}_{j,\omega_{n}}^{\mathrm{tr}}.\nonumber
\end{align}
The contribution of the $\mathbf{\Omega}_0$-term to the action is then given by
(taking $\mathbf{e}_{j}\mathbf{(}0)=\hat{\mathbf{m}}$)%
\begin{equation}
\delta{\mathcal{S}}_{loc}^{\mathrm{dyn}}=-\frac{i}{\beta}\sum_{j}\tanh
(\beta\varphi_{MF})[\mathbf{\Omega}_{j,0}\cdot\mathbf{e}_{j}\mathbf{(}0)]\}].
\end{equation}
It is instructive to write  components of the fluctuation amplitudes  in the following form ($j=(i,\mu)$) 
\begin{align}
&\delta\varphi_{i\mu}^{\alpha}(\tau)   =\sum_{\mathbf{q},\nu=1}^{3N_{s}%
}\operatorname{Re}\{s(\kappa_{\mathbf{q,}\nu})N^{-1/2}e^{-i\mathbf{qR}_{i}%
}u_{\mathbf{q},\nu}^{\mu\alpha}\delta\varphi_{\mathbf{q,}\nu}(\tau)\}\nonumber\\
& =\frac{1}{2}N^{-1/2}\sum_{\mathbf{q},\nu=1}^{3N_{s}}u_{\mathbf{q},\nu}%
^{\mu\alpha}\{s(\kappa_{\mathbf{q,}\nu})e^{-i\mathbf{qR}_{i}}\delta
\varphi_{\mathbf{q,}\nu}(\tau)\\\nonumber & \,\,\,\,\,\,\,\,
+s^{\ast}(\kappa_{\mathbf{q,}\nu}%
)e^{i\mathbf{qR}_{i}}\delta\varphi_{\mathbf{q,}\nu}^{\ast}(\tau)\},
\end{align}
where we assume real-valued, inversion symmetric eigenfunctions $u_{\mathbf{q}%
,\nu}^{\mu\alpha}=(u_{\mathbf{q},\nu}^{\mu\alpha})^{\ast}=u_{-\mathbf{q},\nu
}^{\mu\alpha}$\ . Performing the Fourier transform in time, we get
\begin{align}
\delta\varphi_{i\mu,\omega_{n}}^{\alpha}=&\frac{1}{2}\frac{1}{\sqrt{N}}\sum
_{\mathbf{q},\omega_{n},\nu}u_{\mathbf{q},\nu}^{\mu\alpha}\{s(\kappa
_{\mathbf{q,}\nu})e^{-i\mathbf{qR}_{i}}\delta\varphi_{\mathbf{q,}\omega
_{n},\nu}\nonumber\\&\,\,\,\,\,
+s^{\ast}(\kappa_{\mathbf{q,}\nu})e^{i\mathbf{qR}_{i}}\delta
\varphi_{\mathbf{q,-}\omega_{n}\mathbf{,}\nu}^{\ast}\}.
\end{align}
The contribution of the $\mathbf{\Omega}_0$-term to the local action is then given by
\begin{widetext}
\begin{align}
&\delta{\mathcal{S}}_{loc}^{\mathrm{dyn}} =-\frac{i}{\beta}\frac{(i\beta 
\beta_{c}^{2})}{t}\sum_{\mathbf{q},\omega_{n},\nu,\nu^{\prime}}\omega_{n}%
D_{\mathbf{q}\nu\nu^{\prime}}\Big [ s(\kappa_{\mathbf{q,}\nu})\delta\varphi
_{\mathbf{q,}\omega_{n},\nu}s(\kappa_{-\mathbf{q,}\nu^{\prime}})\delta
\varphi_{-\mathbf{q,}\omega_{n},\nu^{\prime}}+\nonumber\\&
  s^{\ast}(\kappa_{\mathbf{q,}\nu})\delta\varphi_{\mathbf{q,-}\omega_{n}%
,\nu}^{\ast}s^{\ast}(\kappa_{-\mathbf{q,}\nu^{\prime}})\delta\varphi
_{-\mathbf{q,-}\omega_{n},\nu^{\prime}}^{\ast}+
s(\kappa_{\mathbf{q,}\nu
})\delta\varphi_{\mathbf{q,}\omega_{n},\nu}s^{\ast}(\kappa_{\mathbf{q,}%
\nu^{\prime}})\delta\varphi_{\mathbf{q,-}\omega_{n},\nu^{\prime}}^{\ast}
  +s^{\ast}(\kappa_{\mathbf{q,}\nu})\delta\varphi_{\mathbf{q,-}\omega_{n}%
,\nu}^{\ast}s(\kappa_{\mathbf{q,}\nu^{\prime}})\delta\varphi_{\mathbf{q,}%
\omega_{n},\nu^{\prime}}\Big]\nonumber\\
 & =\frac{\beta_{c}^{2}}{t}\sum_{\mathbf{q},\omega_{n},\nu,\nu^{\prime}}%
\omega_{n}D_{\mathbf{q}\nu\nu^{\prime}}\delta\varphi_{\mathbf{q},-\omega
_{n},\nu}^{(+)}\delta\varphi_{\mathbf{q},\omega_{n},\nu^{\prime}}^{(+)\ast
},\\\nonumber
&D_{\mathbf{q}\nu\nu^{\prime}}=\sum_{\mu}\sum_{\alpha_{1},\alpha_{2},\alpha
_{3}}\sum_{\alpha_{2}^{\prime},\alpha_{3}^{\prime}}m_{\alpha_{1}}%
\epsilon_{\alpha_{1}\alpha_{2}\alpha_{3}}P_{\alpha_{2}\alpha_{2}^{\prime}%
}P_{\alpha_{3}\alpha_{3}^{\prime}}u_{\mathbf{q}\nu}^{\mu\alpha_{2}^{\prime}%
}u_{\mathbf{q}\nu^{\prime}}^{\mu\alpha_{3}^{\prime}}.
\end{align}
\end{widetext}

The dynamic fluctuation expression for the exchange interaction term is given
by
\begin{align}
\delta S_{\kappa}  &  =\frac{1}{\beta}\int_{0}^{\beta}d\tau\sum_{\mathbf{q},\nu
}|\kappa_{\mathbf{q,}\nu}|^{-1}\delta\varphi_{\mathbf{q,}\nu}(\tau
)\delta\varphi_{\mathbf{q,}\nu}^{\ast}(\tau)=\nonumber\\
&  =\sum_{\mathbf{q},\omega_{n}\neq0,\nu}|\kappa_{\mathbf{q,}\nu}|^{-1}%
\delta\varphi_{\mathbf{q,}\omega_{n},\nu}^{(+)}\delta\varphi_{\mathbf{q,-}%
\omega_{n},\nu}^{(+)\ast}.%
\end{align}
 Adding the two contributions we find%
\begin{equation}
\delta{\mathcal{S}}^{\mathrm{dyn}}=\beta^{-1}\sum_{\mathbf{q,}\omega_{n}%
\neq0;\nu,\nu^{\prime}}A_{\mathbf{q,}\omega_{n};\nu\nu^{\prime}}%
^{\mathrm{dyn}}\delta\varphi_{\mathbf{q},0,\nu}^{(+)\mathrm{\ast}}%
\delta\varphi_{\mathbf{q},0,\nu^{\prime}}^{(+)},
\end{equation}
where $A_{\mathbf{q,}\omega_{n};\nu\nu^{\prime}}^{\mathrm{dyn}}$ is given by\
\begin{equation}
A_{\mathbf{q,}\omega_{n};\nu\nu^{\prime}}^{\mathrm{dyn}}=\beta\lbrack
|\kappa_{\mathbf{q,}\nu}|^{-1}\delta_{\nu\nu^{\prime}}+\beta_{c}^{2}%
t^{-1}\omega_{n}D_{\mathbf{q}\nu\nu^{\prime}}].
\label{Adyn-general}
\end{equation}
Now one may perform the integration over the fluctuation amplitudes resulting
in
\[
Z^{\mathrm{dyn}}=Z^{\mathrm{MF}}\exp\left[  -\beta\delta{\mathcal{F}%
}^{\mathrm{dyn}}\right] ,
\]
which gives the free energy contribution to be equal to
\begin{equation}
\delta{\mathcal{F}}^{\mathrm{dyn}}=\frac{1}{2\beta}\sum_{\mathbf{q,}\omega_{n}\neq
0}\ln\det\{A_{\mathbf{q,}\omega_{n};\nu\nu^{\prime}}^{\mathrm{dyn}}\}.
\end{equation}

In summary, in the approximation in which the corrections to the free energy
come predominantly from Gaussian fluctuations, the partition function is found
to be
\[
Z=\frac{Z^{\mathrm{MF}}}{C^{\prime}}\exp\left[  -\beta(\delta\mathcal{F}%
^{\mathrm{stat}}+\delta{\mathcal{F}}^{\mathrm{dyn}})\right] .
\]
  As a sanity check, in the  appendix  we  compute the contribution of dynamic 
  fluctuations to the free energy at low temperature $T<<T_{c}$ and  show that the
contribution of transverse fluctuations from the functional integral
representation recovers the spin wave theory result.

\begin{figure}[ptb]
\label{fig1} \includegraphics[width=0.65\columnwidth]{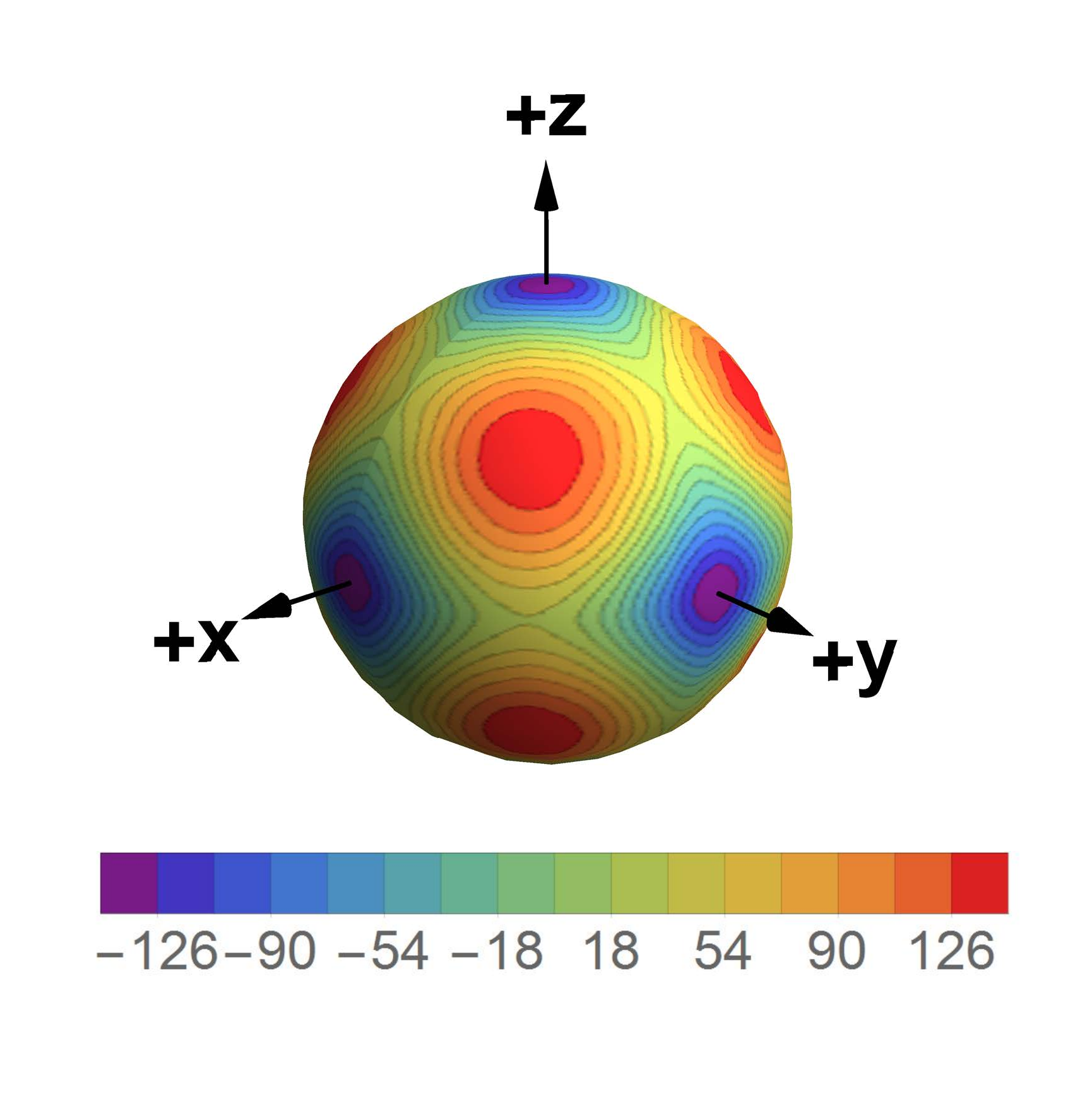}
\caption{(Colors online) (a) The magnitude of the Gaussian contribution to the
free energy $\delta\mathcal{F}(\theta,\phi)$ computed for $J=-1$, $K=-1.1$ and $ \beta=\beta_c+1$
is plotted on the surface of the unit sphere. The minima of the free energy
are shown by deep blue color. The preferred directions of the magnetization
are along the cubic axes. The energy scale is shown in units of $J$.}%
\end{figure}

\section{Lifting mean-field degeneracies in quantum Heisenberg-compass spin
model}
Here,   we compute the contribution of Gaussian fluctuations to the free energy   in the quantum Heisenberg-compass model on the cubic lattice. This model  is  one of the simplest  models described by  Eq.
(\ref{eq:ham}), in which the interaction matrix has only diagonal elements:
\[
J_{j,j^{\prime}}^{\alpha\alpha}=\frac{1}{2}\delta_{j^{\prime},j+\tau_{\mu}%
}\left[  J+K\delta_{\alpha,\mu}\right]  ,
\]
where $\tau_{\mu}=\pm x,\pm y,\pm z$ labels nearest neighbor bonds. The
eigenvalues of the exchange operator are  given by
\[
\kappa_{\mathbf{q},\nu}=\sum_{\alpha}(J+K\delta_{\alpha,\nu})\cos q_{\alpha}.
\]
The three eigenvectors $\mathbf{u}_{\mathbf{q}\nu}$\ point along the three
cubic axes, such that the components are $u_{\mathbf{q}\nu}^{\alpha}%
=\delta_{\nu,\alpha}$, $\alpha=x,y,z$. Provided $J<0$ and $K<|J|$ the
ferromagnetic mean field solution $\varphi_{\mathrm{MF}}$ is given by the
solution of the non-linear equation $2|\kappa_{0}|^{-1}\varphi_{\mathrm{MF}%
}=\tanh(\beta\varphi_{\mathrm{MF}})$.

The fluctuation contribution is described by a $3\times3$-matrix
\begin{equation}
A_{\mathbf{q,}\omega_{n},\nu\nu^{\prime}}=\delta_{\omega_{n},0}%
A_{\mathbf{q},\nu\nu^{\prime}}^{stat}+(1-\delta_{\omega_{n},0})A_{\mathbf{q,}%
\omega_{n},\nu\nu^{\prime}}^{dyn},
\end{equation}
whose matrix elements  can be easily
obtained for arbitrary orientation of magnetization, $\hat{\mathbf{m}}=(\sin
\theta\cos\phi,\sin\theta\sin\phi,\cos\theta)$.   Using Eq.(\ref{A_stat}), we get the following expression for  the static contribuion:%
\begin{equation}
A_{\mathbf{q},\nu\nu^{\prime}}^{stat}=\left(
\begin{array}
[c]{ccc}%
d_{\mathbf{q},x} & g_{x,y} & g_{x,z}\\
g_{y,x} & d_{\mathbf{q},y} & g_{y,z}\\
g_{z,x} & g_{z,y} & d_{\mathbf{q},z}%
\end{array}
\right),  \label{cubicmatrixstat}%
\end{equation}
where  $d_{\mathbf{q},\nu}=|\beta\kappa_{\mathbf{q},\nu}^{-1}%
|-b_{m}m_{\nu}^{2}-{b}_{tr}$, $g_{\nu\nu^{\prime}}=-b_{m}m_{\nu}m_{\nu
^{\prime}}$,  $\kappa_{\mathbf{q},\nu}^{-1}=1/(3J\gamma_{\mathbf{q}}+K\cos
q_{\nu}),$ $\gamma_{\mathbf{q}}=\frac{1}{3}\sum_{\beta}\cos q_{\beta}$,
$b_{m}=\frac{1}{2}(1-t^{2})$, ${b}_{tr}=\frac{1}{2}\beta_{c}\beta$ . We
remind that $t=\tanh(\beta\varphi_{\mathrm{MF}})$  is a dimensionless measure of magnetization.

The dynamical  matrix  is defined by  Eq.\ref{Adyn-general}, which for the cubic geometry  simplifies  to the following expression:
\begin{align}
A_{\mathbf{q,}\omega_{n},\nu\nu^{\prime}}^{dyn}=&\beta^{-1}[|\kappa
_{\mathbf{q,}\nu}|^{-1}\delta_{\nu,\nu^{\prime}}\\\nonumber
+&\omega_{n}\beta_{c}^{2}%
t^{-1}\sum_{\alpha_{1},\alpha_{2},\alpha_{3}}m_{\alpha_{1}}\epsilon_{\alpha_{1}\alpha_{2}\alpha_{3}}P_{\alpha_{2}\nu}P_{\alpha_{3}\nu^{\prime}}\label{Adyncubic}
\end{align}

The matrix $A_{\mathbf{q,}\omega_{n},\nu\nu^{\prime}}$ may be diagonalized for
fixed $\mathbf{q,}\omega_{n}$. Its eigenvalues $\lambda_{\nu,\mathbf{q}%
,\omega_{n}}=\lambda_{\nu,\mathbf{q}%
,\omega_{n}} (\theta,\phi)$ have a rather complex dependence on angles $\theta$ and $\phi$, implying an angular dependent profile of the fluctuation free energy
$\delta\mathcal{F}=\delta\mathcal{F}(\theta,\phi)$.
After integration over fluctuations we obtain%
\begin{align}
Z  &  =\frac{Z^{\mathrm{MF}}}{C^{\prime}}\exp\left[  -\beta\delta
\mathcal{F}\right] ,\\\nonumber
\delta\mathcal{F}  &  \mathcal{=}\frac{1}{2\beta}\sum_{\mathbf{q},\omega_{n},\nu}%
\ln\lambda_{\nu,\mathbf{q},\omega_{n}}+const\,.
\end{align}
 In performing the summation over the Matsubara
frequencies, we need to regularize the expression by subtracting a term
$\ln[\omega_{n}\beta_{c}^{2}/\beta t]$ from $\ln\lambda_{\nu,\mathbf{q}%
,\omega_{n}}$, which will guarantee convergence of the $\omega_{n}$
summation. The subtracted term corresponds to the fluctuation free energy at
the transition point.

In Fig.1, we show the angular dependence of $\delta\mathcal{F}(\theta,\phi)$
computed for representative parameters $J=-1$ and $K=-1.1$. The magnitude of
$\delta\mathcal{F}(\theta,\phi)$ is presented as a color-coded plot on the
unit sphere, where the minima and maxima of the free energy are shown by deep
blue and red color, correspondingly. We see that the minima of $\delta
\mathcal{F}(\theta,\phi)$ are achieved when the magnetization is directed
along one of the cubic axes. This finding clearly shows that while mean field free
energy is isotropic, the fluctuation free energy depends upon the direction of
the order parameter, indicating that the contribution of fluctuations to the
free energy removes the degeneracy of the equilibrium state found on the mean
field level.

\section{Conclusion}

In summary, in this paper we elaborate on  a method for calculating the free energy
of quantum spin systems using functional integral techniques. 
We employ  a powerful formal technique known as the Hubbard-Stratonovich  transformation to map an interacting quantum spin system
into  a collection of "single spin"-systems coupled to a fictitious fluctuating magnetic field.
This method  is very general and can be applied to any  biquadratic  quantum spin model.
Indeed, the Hubbard-Stratonovich transformation applied to isotropic Heisenberg systems in the low temperature limit has been considered before,\cite{angelucci89,angelucci91,angelucci92}  but has not been applied, as far as we know, to calculate the free energy of anisotropic quantum spin systems. 
   In this work, we present  a microscopic derivation of the path-integral representation
of the quantum-spin-system partition function for a particular class of  quantum spin models with anisotropic bond-directional spin interactions. We determine the contribution of Gaussian fluctuations to the free energy at all temperatures in the ordered phase.  Our analysis shows explicitly
that the fluctuation free energy   has  a complex angular
dependence, thus breaking the rotational degeneracy of the mean-field ground state.

  We believe that the proposed method holds good promise to understand
directional ordering in systems with anisotropic bilinear interactions, which
are common in SOC systems. In these systems, the high degeneracy of the
mean-field solution is lifted by the anisotropy of the spin-spin interaction,
such that the spontaneous magnetization is pinned along certain preferred
directions. The latter may change with temperature. 

For illustration,  we apply the above
analysis to the quantum Heisenberg-compass spin model and show that the
direction of the order parameter in spin space is selected by fluctuations and
is determined by the competition between Heisenberg and compass terms. For the
range of parameters for which the ferromagnetic state is the ground state, the
Gaussian fluctuations select the cubic axes as directions of the magnetization.

\section{Acknowledgements}

We thank Ioannis Rousochatzakis for useful discussions. P.W. thanks the
Department of Physics at the University of Wisconsin-Madison for hospitality
during several stays 2011-2014 as a visiting professor. P.W. also acknowledges
partial support by an ICAM senior fellowship. Part of this work was performed
during the summer of 2015 at the Aspen Center for Physics, which is supported
by NSF Grant No. PHY-1066293. N.P. and Y.S. acknowledge the support from NSF
DMR-1511768 Grant. N.P. acknowledges the hospitality of KITP and partial
support by the National Science Foundation under Grant No. NSF PHY11-25915.

\bigskip
\appendix

\section{Isotropic Heisenberg model}

Here we calculate the contribution of dynamic fluctuations to the free energy
at low temperature $T<<T_{c}$. It is known that the leading contribution comes
from spin wave excitations. The purpose of this appendix is to show that the
contribution of transverse fluctuations from the functional integral
representation recovers the spin wave theory result.

As a simple example we calculate the contribution of dynamic fluctuations for
the case $\kappa_{\mathbf{q,}\nu}=3J\gamma_{q}$, where $\gamma_{q}=\frac{1}%
{3}\sum_{\alpha}\cos q_{\alpha}$ and $J<0$. 
The spin wave excitation energy in our representation  is given by 
\begin{equation}
\omega_{\mathbf{q}%
}=\frac{t}{\beta_{c}^{2}|\kappa_{\mathbf{q}=0}||\kappa_{\mathbf{q,}\nu}%
|}\Bigl[|\kappa_{\mathbf{q}=0}|-|\kappa_{\mathbf{q,}\nu}|s^{2}(\kappa
_{\mathbf{q}\nu})\Bigr].
\end{equation}
 In the limit $q\ll 1$, it can also be significantly simplified:
\begin{equation}
\omega
_{\mathbf{q}}\approx\frac{3}{4}|J|[1-\gamma_{q}].
\end{equation}
  In this limit, the dynamic fluctuation matrix then takes the form%
\begin{equation}
A_{\mathbf{q,}\omega_{n};\nu,\nu^{\prime}}^{\mathrm{dyn}}=
\frac{\beta
_{c}^{2}}{\beta t}
\left(
\begin{array}
[c]{cc}%
\omega_{\mathbf{q}} & \omega_{n}\\
-\omega_{n} & \omega_{\mathbf{q}}%
\end{array}
\right).
\end{equation}
Its determinant is equal to
\begin{equation}\det\{A_{\mathbf{q,}\omega_{n};\nu,\nu^{\prime}}^{\mathrm{dyn}}\}=\left(\frac{\beta
_{c}^{2}}{\beta t}\right)^2\left [\omega_{n}^{2}+\omega_{\mathbf{q}}^{2}\right].
\end{equation}
Recalling that
 the transverse  fluctuation free energy is  given by
\[ 
\delta{\mathcal{F}}^{\mathrm{tr}}=\frac{1}{2\beta}\sum_{\mathbf{q,}\omega_{n}\neq
0}\ln\det\{A_{\mathbf{q,}\omega_{n};\nu\nu^{\prime}}^{\mathrm{dyn}}\},
\]
 the  contribution to the partition function is
found  to be
\begin{align}
&Z^{\mathrm{tr}}=\exp[-\beta\delta{\mathcal{F}}^{\mathrm{tr}}]  =
\exp[-\frac{1}{2}\sum_{\mathbf{q,}\omega_{n}\neq
0}\ln\det\{A_{\mathbf{q,}\omega_{n};\nu\nu^{\prime}}^{\mathrm{dyn}}\}]\nonumber
\\&=\exp[-\frac{1}{2}\sum_{\mathbf{q},\omega_{n}\neq
0}\{\ln(\beta^{-1}\beta_{c}^{2}t^{-1}%
)^{2}+\ln[\omega_{n}^{2}+\omega_{\mathbf{q}}^{2}]\}.
\end{align}
The first term in the curly  brackets give simple constant renormalization.  The  summation  over Matsubara frequencies in the second term gives%
\begin{align}
X_{\mathbf{q}}  &  =\operatorname{Re}\frac{1}{2}\sum_{\omega_{n}}\ln[\omega_{n}%
^{2}+\omega_{\mathbf{q}}^{2}]=\operatorname{Re}\frac{1}{2}\sum_{\omega_{n}}\ln[(i\omega_{n})^{2}%
-\omega_{\mathbf{q}}^{2}]\\
&  =\frac{1}{2}\beta\omega_{\mathbf{q}}+\ln\left[1-\exp(-\beta\omega_{\mathbf{q}})\right ],
\end{align}
which leads to the free energy contribution%
\begin{equation}
\delta{\mathcal{F}}^{\mathrm{tr}}=\frac{1}{2}\sum_{\mathbf{q}}\{\omega_{\mathbf{q}%
}+2\beta^{-1}\ln[1-\exp(-\beta\omega_{\mathbf{q}})]\}+const.
\end{equation}
To get this result we differentiate $X_{\mathbf{q}}$ with respect to
$\omega_{\mathbf{q}}$%
\begin{align}
\frac{\partial}{\partial\omega_{\mathbf{q}}}X_{\mathbf{q}}  &  =\frac{1}{2}\sum
_{\omega_{n}}[\frac{1}{i\omega_{n}+\omega_{\mathbf{q}}}-\frac{1}{i\omega
_{n}-\omega_{\mathbf{q}}}]\\
&  =\frac{1}{2}\beta\lbrack-n_B(-\omega_{\mathbf{q}})+n_B(\omega_{\mathbf{q}})]
\end{align}
where $n_B(\omega_{\mathbf{q}})=[\exp(\beta\omega_{\mathbf{q}})-1]^{-1}$ is the
Bose distribution function. 

The term $\sum_{\mathbf{q}}\omega_{\mathbf{q}}$ is the zero
point fluctuation contribution of the two transverse modes to the ground state
energy (note that $\omega_{\mathbf{q}}$ is independent of $\nu$ for the
isotropic model considered). There must be an additional constant contribution
$\delta\omega_{0}$ to the ground state energy, which is not completely captured by the Gaussian fluctuation contribution, such that $\sum_{\mathbf{q}}\omega_{\mathbf{q}}+$
$\delta\omega_{0}\propto\sum_{\mathbf{q}}
\gamma_{\mathbf{q}}$, which sums to zero. Recall that
for the isotropic model the ground state is identical to the mean field ground
state, such that the fluctuation contribution to the ground state energy vanishes.
The  fluctuation contribution to the internal energy  is  then given 
\begin{equation}
\delta U^{\mathrm{dyn}}=\delta{\mathcal{F}}^{\mathrm{dyn}}+\beta\frac
{\partial\delta{\mathcal{F}}^{\mathrm{dyn}}}{\partial\beta}=\sum_{\mathbf{q}%
}\omega_{\mathbf{q}}\{\frac{1}{2}+n_B(\omega_{\mathbf{q}})\}
\end{equation}
This is identical with the standard result of spin wave theory, except that
$\omega_{\mathbf{q}}$ differs from the spin wave result at higher $q$.\ At low
temperatures $\delta F^{\mathrm{dyn}}$ provides the leading contribution to
the thermodynamic quantities, e.g. $\delta U^{\mathrm{dyn}}\propto T^{5/2}$,
whereas the longitudinal fluctuations contribute an exponentially small term.
As $\omega_{\mathbf{q}}%
=\frac{1}{2}|J|q^{2}+O(q^{4}),$  the leading low temperature behavior of $\delta U^{\mathrm{dyn}}$ agrees exactly with the conventional spin wave result.


\begin{thebibliography}{99}                                                                                               %

\bibitem {khaliullin05}G. Khaliullin, Prog. Theor. Phys. Suppl. \textbf{160}%
,155 (2005).

\bibitem {jackeli09}G. Jackeli and G. Khaliullin, Phys. Rev. Lett.
\textbf{102}, 017205 (2009).

\bibitem {jackeli10}J. Chaloupka, G. Jackeli, and G. Khaliullin, Phys. Rev.
Lett. \textbf{105}, 027204 (2010).

\bibitem {perkins14}N. B. Perkins, Y. Sizyuk and P. W\"{o}lfle, Phys. Rev. B
\textbf{89}, 035143 (2014).

\bibitem {sizyuk14}Y. Sizyuk, C. Price, P. W\"{o}lfle, and N. B. Perkins,
Phys. Rev. B \textbf{90}, 155126 (2014).

\bibitem {rau14}J. G. Rau, E. Kin-Ho Lee, H.-Y. Kee, Phys. Rev. Lett.
\textbf{112}, 077204 (2014).

\bibitem {kimchi14}I. Kimchi and A. Vishwanath, Phys. Rev. B \textbf{89},
014414 (2014).

\bibitem {nussinov15}Z. Nussinov, J. van den Brink, Rev. Mod. Phys.
\textbf{87} 1 (2015).

\bibitem {balents14}H. Ishizuka and  L. Balents, Phys. Rev. B \textbf{90},
184422 (2014).

\bibitem {chaloupka15} J. Chaloupka and G. Khaliullin, Phys. Rev. B
\textbf{92}, 024413 (2015).
\bibitem {chaloupka16} J. Chaloupka and G. Khaliullin, Phys. Rev. B
\textbf{94}, 064435 (2016).

\bibitem {ioannis15}I. Rousochatzakis, J. Reuther, R. Thomale, S. Rachel and
N. B. Perkins, PRX \textbf{5}, 041035 (2015).
\bibitem {ioannis16}I. Rousochatzakis and
N. B. Perkins,  arXiv:1610.08463. 

\bibitem {trebst15}Michael Becker, Maria Hermanns, Bela Bauer, Markus Garst,
Simon Trebst, Phys. Rev. B \textbf{91}, 155135 (2015).

\bibitem {jackeli15}G. Jackeli and A. Avella, Phys. Rev. B \textbf{92}, 184416 (2015).



\bibitem{rau16}J. G. Rau, E. Kin-Ho Lee, H.-Y. Kee, Annual Review of Condensed Matter Physics  {\bf 7}, 195 (2016). 


\bibitem {kitaev06}A. Kitaev, Ann. Phys. \textbf{321}, 2 (2006).

\bibitem {choi12}S. K. Choi, R. Coldea, A. N. Kolmogorov, T. Lancaster, I. I.
Mazin, S. J. Blundell, P. G. Radaelli, Yogesh Singh, P. Gegenwart, K. R. Choi,
S.-W. Cheong, P. J. Baker, C. Stock, and J. Taylor, Phys. Rev. Lett.
\textbf{108}, 127204 (2012).

\bibitem {chun15}S. H. Chun, J.-W. Kim, J. Kim, H. Zheng, C. C. Stoumpos, C.
D. Malliakas, J. F. Mitchell, Kavita Mehlawat, Yogesh Singh, Y. Choi, T. Gog,
A. Al-Zein, M. Moretti Sala, M. Krisch, J. Chaloupka, G. Jackeli, G.
Khaliullin, B. J. Kim, Nature Physics \textbf{11}, 462 (2015).

\bibitem {banerjee2015}A. Banerjee, C.A. Bridges, J-Q. Yan, A.A. Aczel, L. Li,
M.B. Stone, G.E. Granroth, M.D. Lumsden, Y. Yiu, J. Knolle, D.L. Kovrizhin, S.
Bhattacharjee, R. Moessner, D.A. Tennant, D.G. Mandrus, S.E. Nagler, Nature materials {\bf 15}, 733 (2016).

\bibitem {johnston15}R. D. Johnson, S. Williams, A. A. Haghighirad, J.
Singleton, V. Zapf, P. Manuel, I. I. Mazin, Y. Li, H. O. Jeschke, R. Valenti,
R. Coldea, Phys. Rev. B {\bf 92}, 235119 (2015).

\bibitem {biffin14-1}A. Biffin, R. D. Johnson, S. Choi, F. Freund, S. Manni,
A. Bombardi, P. Manuel, P. Gegenwart, and R. Coldea, Phys. Rev. B \textbf{90},
205116 (2014).

\bibitem {biffin14-2}A. Biffin, R.D. Johnson, I. Kimchi, R. Morris, A.
Bombardi, J.G. Analytis, A. Vishwanath, and R. Coldea, Phys. Rev. Lett.
\textbf{113} 197201 (2014).

\bibitem {tomo15}T. Takayama, A. Kato, R. Dinnebier, J. Nuss, H. Kono, L.S.I.
Veiga, G. Fabbris, D. Haskel, and H. Takagi, Phys. Rev. Lett. \textbf{114},
077202 (2015).


\bibitem {sizyuk16} Y. Sizyuk, P. W\"{o}lfle and N. B. Perkins, Phys. Rev. B
\textbf{94}, 085109 (2016).



\bibitem{villain80} J. Villain, R. Bidaux, J.-P. Carton, and R. Conte, J. Phys. France {\bf 41},
1263 (1980).
\bibitem{shender82} E. F. Shender, Sov. Phys. JETP {\bf 56}, 178 (1982).

\bibitem{chubukov92} A. Chubukov, Phys. Rev. Lett. {\bf 69}, 832  (1992).
\bibitem{zhitomirsky12} M. E. Zhitomirsky, M. V. Gvozdikova, P. C. W. Holdsworth, and R. Moessner, Phys. Rev. Lett. {\bf 109}, 077204 (2012).
\bibitem{chern13} G.-W. Chern and R. Moessner Phys. Rev. Lett. {\bf 110}, 077201 (2013).
\bibitem{gingras14} P.A. McClarty, P. Stasiak, and M. J. P. Gingras, Phys. Rev. B {\bf 89}, 024425 (2014).
\bibitem{chernyshev14} A. L. Chernyshev and M. E. Zhitomirsky, Phys. Rev. Lett. {\bf 113}, 237202 (2014).
\bibitem{chernyshev15} A. L. Chernyshev and M. E. Zhitomirsky, Phys. Rev. B {\bf 92}, 144415 (2015).


\bibitem{angelucci89}
A. Angelucci and G. Jug, Int. J. Mod. Phys. B {\bf 3}, 1069 (1989).
\bibitem{angelucci91} A. Angelucci, Phys. Rev. B {\bf 44}, 6849 (1991).
\bibitem{angelucci92} A. Angelucci, Phys. Rev. B {\bf 45}, 5387 (1992).

\bibitem {Negele1998}J. W. Negele and H. Orland, Quantum Many Particle Systems
Westview Press, Boulder (1998).

\bibitem {Fradkin2013}E. Fradkin, Field Theories of Condensed Matter Physics
Cambridge University Press, Cambridge, UK (2013).

\bibitem {hubb59}J. Hubbard, Phys. Rev. Lett. \textbf{3}, 77 (1959).

\bibitem {strat58}R. L. Stratonovich, Sov. Phys. Dokl. \textbf{2}, 416 (1958).

\bibitem {sizyuk15}Y. Sizyuk, N. B. Perkins and P. W\"{o}lfle, Phys. Rev. B
\textbf{92}, 155131 (2015).


\bibitem {trott59}H. Trotter, Proc. Am. Math. Soc. \textbf{10}, 545 (1959).

\end{thebibliography}
\end{document}